\documentclass{mem}
\usepackage{natbib}\usepackage{txfonts}\usepackage{balance}
\usepackage{graphicx}
\usepackage[a4paper]{hyperref}
\idline{75}{1}
\begin{document}
\def\teff{$T\rm_{eff }$}
\def\kms{$\mathrm {km s}^{-1}$}
\def\hii{H~{\sc ii}}
\def\cbeta{c$\beta$}
\def\hb{H$\beta$}
\def\ha{H$\alpha$}
\def\te{$t_{e}$}
\def\deg{$^\circ$}
\title{
Metals in M33
}
\subtitle{Radial gradients and 2D maps}

\author{
Laura  \,Magrini\inst{1} 
%\and B. \, Rabbit\inst{1,2}
          }

%  \offprints{L. Magrini}

\institute{
Istituto Nazionale di Astrofisica --
Osservatorio Astrofisico di Arcetri, Largo E. Fermi, 5,
I-50125 Firenze, Italy
%\and
%Forest  University, Department of Astronomy,
%25 Long Street, 255255,
%Somewhere, Elsewhere 
\email{laura@arcetri.astro.it}
}

\authorrunning{Magrini }

\titlerunning{Metals in M33}

\abstract{The metal content of the spiral galaxy M33 is analyzed through spectroscopic observations 
of its emission-line populations, \hii\ regions and Planetary Nebulae (PNe): their abundance 
gradients are identical within the errors. 
The 2D metallicity map is presented, finding an off-center peak, located in the southern arm. 
A chemical evolution model of M33 described by Magrini et al. (2007a) is updated 
at the  light of recent results on the Schmidt law and the metallicity gradients. 
}
\maketitle{}

\section{Introduction}
M33 (NGC~598) is a spiral galaxy whose closeness (840 kpc, Freedman et al. 1991; optical size 53'
$\times$53', Holmberg 1958) and inclination (i=53 \deg) allow detailed studies of its stellar 
populations and ionized nebulae.
Planetary Nebulae (PNe) and \hii\ regions represent two very different stages in the lifetime of a galaxy, 
and their comparison provides insight on the chemical evolution of the host galaxy. 
PNe are indeed the final ejecta of evolved low- and intermediate-mass stars with mass 
between 1 and 8 M$_{\odot}$, which must have formed between 3$\times$10$^7$ yr and 10 Gyr ago, e.g., 
Maraston \cite{maraston05}, while \hii\ regions belong to a very young population.

One of the most debated questions about the chemical evolution of galaxies is the metallicity gradient 
behaviour with time.  Chemical evolution models predict different temporal behaviors of the metallicity gradient, depending on the assumptions one makes on gas inflow and outflow rates, and the star and cloud formation efficiencies.  Observations are needed to constrain these assumptions, but so far they have been insufficient, especially for the older populations. 
The idea behind the observations leading to our work is to study the chemical and physical properties of a large number of PNe and \hii\ regions in M33, using the same set of observations, the same data reduction and analysis techniques, and identical abundance determination methods, to avoid all biases due to the stellar vs. nebular analysis. The aim is to derive abundances of the $\alpha$-elements for as many PNe 
and \hii\ regions as possible, and to study the variation of the metallicity gradient and the average 
abundances in M33.

In this paper, several aspects of the study of M33 will be described. 
In Section \ref{sect_mmt} the observations of a sample of PNe and \hii\ regions with MMT will be presented. 
These data, together with a large literature data-set, allow us to recompute the PN and \hii\ region 
metallicity radial gradients.
In Section \ref{sect_distr} the spatial distribution of the metallicity will be shown. 
The off-center of the metallicity maps obtained from \hii\ regions and from PNe  will be discussed.
In Section \ref{sect_model} the chemical evolution model of M33 (Magrini et al. 2007a, hereafter M07) will be revised, 
including a star formation process, parametrized by a Schmidt law, consistent with the observations. 
The consequences, in particular the 
time evolution of the metallicity gradient, will be analyzed. 
Finally, our conclusions and summary will be given in Section \ref{sect_conclu}.

\section{The MMT observations}
\label{sect_mmt}
In semester 2007B we observed $\sim$150 ionized nebulae in M33 in multi-object spectroscopic mode.
We used the Hectospec fiber-fed spectrograph (Fabricant et al. 2005) on the Multi Mirror Telescope 
(MMT), with a 270 mm$^{-1}$ grating at a dispersion of 1.2 \AA\ pixel$^{-1}$. 
The instrument deploys 300 fibers over a 1\deg\ diameter field of view; the fiber diameter is 1.5\arcsec\ (6 pc using a distance of 840 kpc to M33).  
We obtained spectra of 102 PNe and 48 HII regions with resulting total spectral coverage from ~3600 to 9100 \AA. 
The details on the data reduction and analysis are in Magrini et al. (2009, hereafter M09). 
Most of the PNe observed in M33 belong to its disk, and they are non-Type I, implying a population 
mainly composed by PNe with old progenitors, i.e. M$<$3 M$_{\odot}$, and ages $>$0.3 Gyr.

The MMT observations allowed us to obtain a good determination of the O/H gradient (together with 
Ne/H, S/H, and Ar/H) both from \hii\ regions and PNe. In the case of \hii\ regions, our cumulative sample includes: 
i)  \hii\ regions by Magrini et al. \cite{magrini07b} which comprises their own determinations and all previous abundance determinations  with available \te, recomputed uniformly; ii) the sample by  Rosolowsky \& Simon 
(2008); iii) the new MMT sample. For PNe, we use the sample of PNe presented by M09. 
The resulting gradients are the following, for PNe and \hii\ regions respectively, 12 + {\rm log(O/H)} =  
$$   -0.031 (\pm 0.013) ~  {\rm R_{GC}} + 8.44 (\pm
0.06),\eqno(1)$$ 
$$   -0.032 (\pm 0.009) ~  {\rm R_{GC}} + 8.42 (\pm
0.04),\eqno(2)$$ where R$_{GC}$ is the galactocentric distance in kpc. 
The two gradient are identical within the errors, both in the their slopes and central values. 
\section{The metallicity in M33 and its evolution}
\label{sect_distr}

The large amount of chemical abundance data from \hii\ regions and PNe collected 
to date in M33 allow us to reconstruct  its metallicity map.
In this section, both the 2D  and the radial distribution are analyzed
taking advantages also of all previous abundance determinations with measured \te\ and of our new
results.

\subsection{The 2D distribution of metals}

The usual way to study the metallicity distribution in disk galaxies is 
to average it azimuthally, assuming that: i)  the centre of the galaxy coincides
with the peak of the metallicity distribution; ii) at a given radius, the metallicity is the 
same in each side of the galaxy. 
The large number of metallicity measurements in M33, both from \hii\ regions and from PNe, 
allow us to reconstruct not only their radial gradient, but also their spatial distribution 
projected onto the disk.  
In Figure 1, the two-dimensional metallicity distributions for M33 from \hii\ regions and 
from PNe are shown. 
The highest metallicity \hii\ regions and PNe are not located at the center of the
galaxy, but rather lie at a radius of 1-2 kpc, in the southern arm.  
%Also in the case of PNe, 
%most of the metal rich PNe are located in the souther part of M33. 
Simon \& Rosolowsky (2008) noticed this behavior for \hii\ regions and suggested as 
explanation that the material enriched by the most recent generation
of star formation in the arm has not yet been azimuthally  mixed through the
galaxy.  Whilst it might be true for \hii\ regions, it cannot be the reason for the off-center 
metallicity distribution of PNe since they belong to an older population. 
Colin \& Athanassoula \cite{colin81} noticed several evidences of asymmetries in the inner regions 
of M33, such as the distribution of HI atomic gas, of \hii\ regions, of  high luminosity stars. They
proposed a kinematical model with a displaced bulge having a retrograde motion around the center. 
Also detailed analysis of the innermost regions of M33 by Corbelli \& Walterbos \cite{corbelli07} 
found possible asymmetries in the stellar and gas velocity pattern which might  be related 
to the displacement of a small bulge.
The off-center metallicity distribution might thus be related to the lack of a dominant gravitational 
source in the center of this galaxy with a consequent motion at different epochs of the peak of the highest star 
formation region around the M33 visual center.  
 
 \begin{figure}
  \resizebox{\hsize}{!}{\includegraphics[clip=true,angle=270]{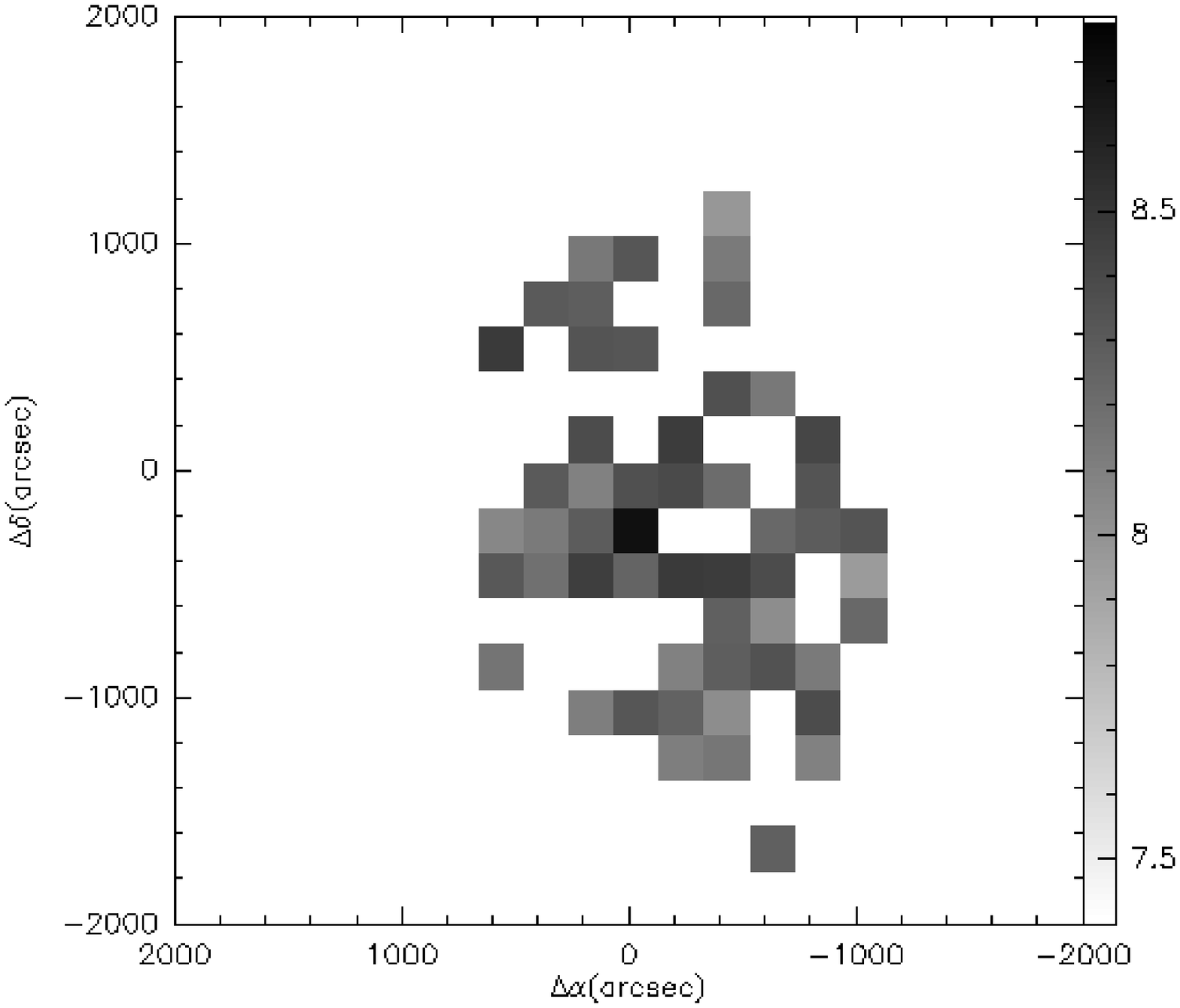}}
  \resizebox{\hsize}{!}{\includegraphics[clip=true,angle=270]{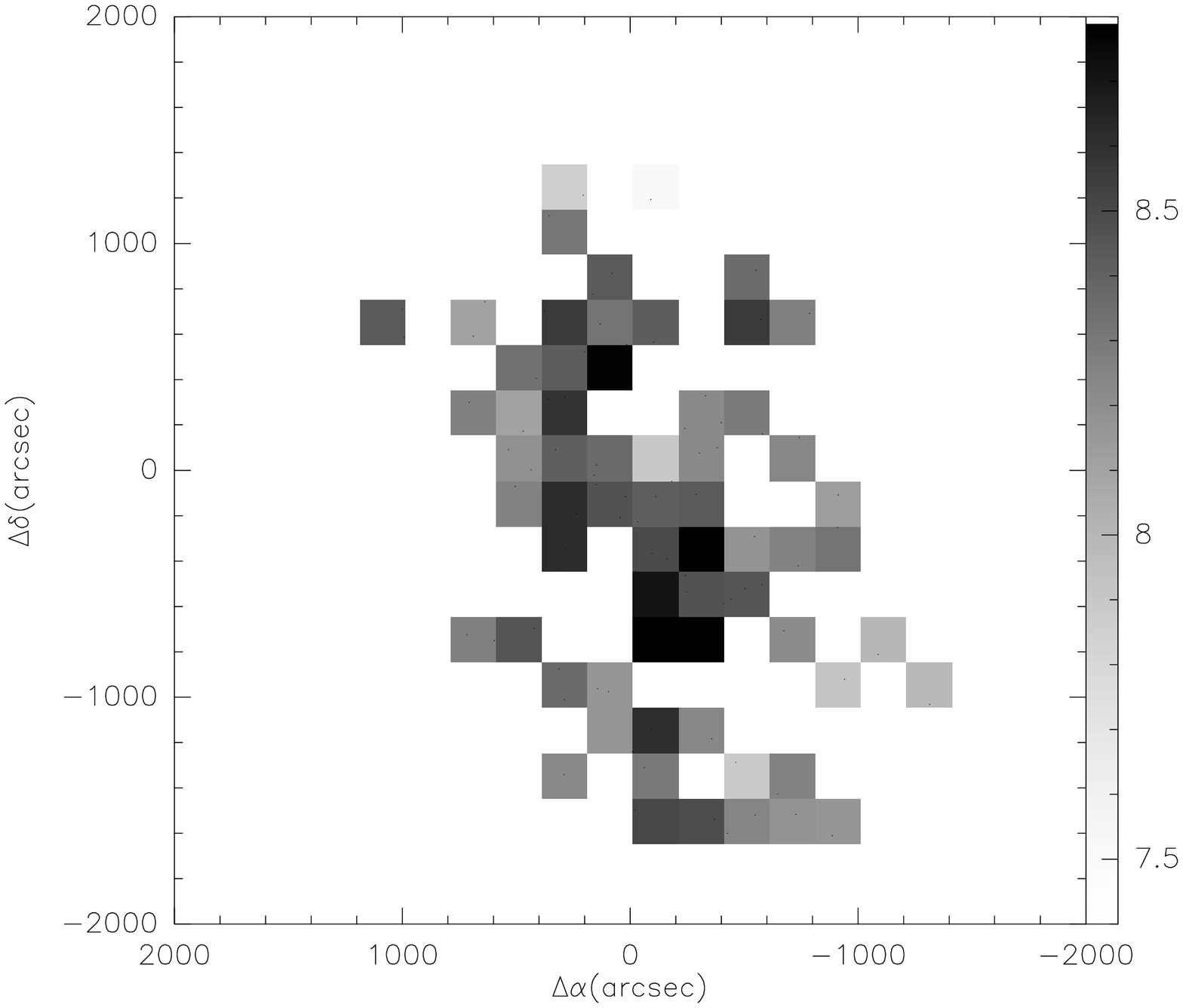}}
  \caption{\footnotesize{The metallicity maps: PNe (top) and \hii\ regions (bottom). 
  The 12 + log(O/H) scale is shown on the y axe. The centre of M33 corresponds to the 0,0 position.      }}
  \label{oxy}%
   \end{figure}

\subsection{The time-evolution of the abundance gradient}
\label{sect_model}

M07 built a chemical evolution model of M33, called {\em accretion} model,  
able to reproduce its main features, 
including the radial trends at  present-time of molecular gas, atomic gas,  stars,  SFR, and the 
time evolution of the metallicity gradient using the available  constraints at that time. 
In that model, the disk of M33 was formed by continuous accretion of primordial 
gas from the intergalactic medium. 
However, new observations have been made recently available, rendering necessary 
a revision of the model. 
In particular, the global slope  of the radial O/H gradient  has been confirmed much 
shallower than retained in the past (cf. Rosolowsky \& Simon 2008) 
and its evolution much slower (cf.  M09).
%In this Section the revision of the model is presented: the introduction of a more realistic 
%star formation (SF) law allowed us to reproduce the observed flat gradient and its little evolution 
%with time. 

The general assumption of multiphase chemical evolution models, 
such as the  M33 one (M07), 
is to describe the formation and disruption of diffuse gas, clouds, and stars,   
by means of physical processes, e.g., Ferrini et al. \cite{ferrini92}.  
In particular, the SF is represented 
with two processes: the interaction of molecular clouds  with the 
radiation field of massive stars and the  collisions between two molecular clouds.
The dominant process is due to cloud collisions.  
In this kind of model the relationship between the star formation rate (SFR) and the surface density of gas (molecular or total) is, thus, a by-product of the model, and cannot be assumed 'a priori'.

However, the law relating the surface densities of SF and cold gas is one of the most fundamental laws 
describing the galaxy behaviour. Virtually the entire range of global star formation rates in galaxies 
can be reproduced by a Schmidt power law relation. 
In the particular case of M33, the relation between the SFR, measured from the FUV emission, 
and molecular gas has a well-defined slope (Verley et al., in prep.) corresponding to
\begin{equation}
\Sigma_{SFR} = A  \Sigma^{1.2}_{mol gas}.
\end{equation}
A pure cloud-cloud collision process for the star formation is not able to reproduce it.  
For this reason, we have taken into account other parameterizations of  the SF process. 
In particular, a process which is dominated by cloud collisions close to the center, while in the 
intermediate and peripheral regions is proportional to the fraction of clouds with a 1.5 exponent, 
is able to reproduce the observations. It reproduces  the 
higher cloud surface density in the inner regions, rendering the conversion into star less effective, 
while in the outer regions it takes into account a more efficacious SF.   
The resulting Schmidt law would have an average exponent all over the radial range (or the molecular gas surface density range) of 1.1, thus consistent with the observations. 

In addition, the introduction of the Schmidt law allowed us  to better reproduce 
the O/H gradient and its evolution 
(see Figure 2): a gradient almost flat both at present-time both at the epoch of the formation of the PNe progenitors, 
with a little evolution of its absolute value and slope. 
Note from Figure 2 the steeper gradient predicted by the previous model, where the SF process 
was dominated by cloud-cloud collision all over the radial range. 
  
  \begin{figure}
   \centering
   \resizebox{\hsize}{!}{\includegraphics[clip=true]{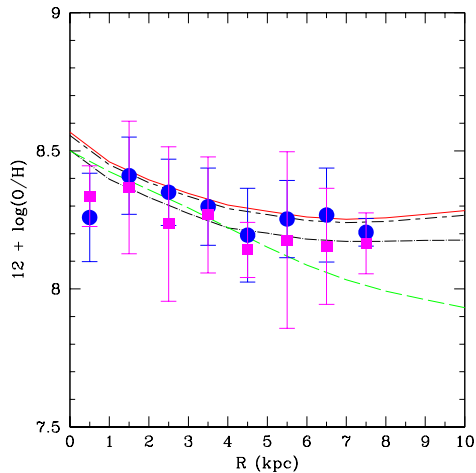}}  
      \caption{\footnotesize{The time-evolution of the O/H radial gradient. Filled symbols  (blue circles and magenta squares) are 
      the oxygen abundances of \hii\ regions and PNe, respectively, averaged in bins 1 kpc wide.
      Model with the Schmidt law: present time (red continuous line), 1 Gyr ago (long-short dashed line), 
      5 Gyr ago (dot-dashed line).  Model with the cloud-cloud collision process (M07): present time (green  dashed line)}. 
       }
              \label{sl}%
    \end{figure}

\section{Summary and conclusions}
\label{sect_conclu}
The chemical evolution of M33 is studied by means of new spectroscopic observations 
of PNe and \hii\ regions.
Their 2D metallicity maps have been found both off-centered, with a peak in the southern
arm, at 1-2 kpc from the center. This might be related to the absence  of a dominant gravitational 
source in the center. 
The slow evolution of the metallicity gradient from the present time to the 
birth of the PNe progenitors is explained with an {\em accretion} model 
where the SF process is driven by the Schmidt law.

\begin{acknowledgements}
I warmly thank Edvige Corbelli, Daniele Galli, 
Letizia Stanghellini, and Eva Villaver for their collaboration in this work. 
\end{acknowledgements}

\bibliographystyle{aa}

\end{document}